\documentclass{article}
\usepackage{spconf,amsmath,graphicx}
\usepackage{booktabs}
\usepackage{color}
\usepackage{comment}

\def \am [#1]{\textcolor{red}{AM: #1}}

\title{A CURATED DATASET of URBAN SCENES FOR AUDIO-VISUAL SCENE ANALYSIS}
\name{Shanshan Wang, Annamaria Mesaros, Toni Heittola, Tuomas Virtanen\thanks{This work was supported in part by the European Research Council under the European Unions H2020 Framework Programme through ERC Grant Agreement 637422 EVERYSOUND. The authors wish to thank CSC-IT Centre of Science Ltd., Finland,  for providing computational resources.}}
\address{Computing Sciences, Tampere University,
Tampere, Finland
\\ \normalsize E-mail: name.surname@tuni.fi}
\begin{document}
%
\maketitle
\begin{abstract}
This paper introduces a curated dataset of urban scenes for audio-visual scene analysis which consists of carefully selected and recorded material. The data was recorded in multiple European cities, using the same equipment, in multiple locations for each scene, and is openly available. We also present a case study for audio-visual scene recognition and show that joint modeling of audio and visual modalities brings significant performance gain compared to state of the art uni-modal systems. Our approach obtained an 84.8\% accuracy compared to 75.8\% for the audio-only and 68.4\% for the video-only equivalent systems. 

\end{abstract}
\section{Introduction}
\label{sec:intro}

Audio-visual analysis of videos has become a popular approach in many applications that use video as their main analysis signal. The joint analysis of the audio and visual modalities brings additional information compared to the visual only data, allowing novel target applications such as visualization of the sources of sound in videos \cite{Arandjelovic_2018_ECCV},  audio-visual alignment for lip-reading \cite{Chung2017}, or audio-visual source separation \cite{Zhao_2018_ECCV}.
In addition, exploiting the audio content can provide novel approaches as a form of cross-modal knowledge distillation for example in concept classification of videos using audio \cite{takahashi2018}, 
multi-view face recognition \cite{sanderson2009multi}, and emotion recognition \cite{zhang2018}, or provide a method for previewing the video and selecting the video segments most useful for video recognition \cite{Gao_2020_CVPR}.

The largest datasets available for audio-visual analysis and classification consist of videos collected from YouTube, with automatically generated label categories    \cite{abu2016youtube, KarpathyCVPR14}. This makes the audio and video content of irregular quality, while the annotations are prone to errors due to the lack of human ratings. For example Sports-1M dataset \cite{KarpathyCVPR14} contains over 1 million video URLs which have been annotated automatically with 487 Sports labels using the YouTube Topics API; YouTube-8M \cite{abu2016youtube} contains 7 million videos,  450 000 hours of video, 3.2 billion audio/visual features, and approximately 5000 labels created by the YouTube video annotation system. In addition, these datasets are not distributed as audio or video data due to copyright: Sports-1M contains video URLs, while YouTube-8M consists of pre-extracted audio and video features. A number of other datasets exist, but they are highly specialized for tasks, focusing for example on human activity \cite{heilbron2015}, action recognition \cite{soomro2012ucf101}, sport types \cite{Gade_2015_ICCV_Workshops}, or emotion \cite{zhang2018}. 

We introduce a carefully planned and recorded audio-visual dataset. In contrast to most of the available ones, this dataset is recorded in real-life environments using the same equipment, which gives it a consistent audio and video quality. In addition, this avoids the problem of confounding variables, where instead of the true analysis target a machine algorithm learns to recognize characteristics of the capture device because of dependencies between the target classes and devices.
Audio and video data were recorded 
such as to not interfere with the spatial perception of sound sources and their corresponding source objects. Moreover, the scenes in which the recordings were performed were planned in advance to constitute homogeneous categories, and planned to have a high diversity, being recorded in 12 large European cities and several locations in each of them. As a first application using this data, we examine the problem of audio-visual scene classification. 
Due to the specific characteristics of the dataset, we expect it to find a variety of uses, for tasks such as audio-visual sound event detection or sound localization by using the relationship between objects and sound sources.

The contributions of this work are the following: 1) we provide a new dataset of audio-visual scenes that is recorded under controlled conditions in comparison with existing ones, while maintaining high variability of content; 2) we introduce audio-visual scene classification (AVSC), using joint modeling of the audio and video content; and 3) we propose a simple architecture for AVSC, based on transfer learning, and show that it outperforms separate domain methods, while being computationally lighter than state-of-the-art approaches.

The paper is organized as follows: Section \ref{sec:data} presents the dataset in detail, Section \ref{sec:avsc} introduces the audio-visual scene classification problem and a proposed system for it. Section \ref{sec:results} presents the results of the classification task and analysis. Finally, Section \ref{sec:concl} presents conclusions and future work.

\section{Audio-visual scenes dataset}
\label{sec:data}

We introduce \textbf{TAU Urban Audio-Visual Scenes 2021}, a dataset recorded in 2018-2019 in 12 large European cities: Amsterdam, Barcelona, Helsinki, Lisbon, London, Lyon, Madrid, Milan, Prague, Paris, Stockholm, and Vienna. It consists of 10 scene classes: airport, shopping mall (indoor), metro station (underground), pedestrian street, public square, street (traffic), traveling by tram, bus and metro (underground), and urban park. Each scene class was defined beforehand based on acoustic and visual characteristics, and suitable locations were selected based on the description.

The data was recorded with four devices recording simultaneously, as explained in \cite{Mesaros2018_DCASE}. TAU Urban Audio-Visual Scenes 2021 contains video data recorded with a GoPro Hero5 Session and the corresponding time-synchronized audio data recorded using a Soundman OKM II Klassik/studio A3 electret binaural in-ear microphone and a Zoom F8 audio recorder with 48~kHz sampling rate and 24 bit resolution.

Recordings were performed in multiple locations for each city and each scene class, i.e. different parks, different streets, different shopping malls; for airport, the locations were selected to be sufficiently far from each other. Locations are identified by a running ID number, used for identifying all data from the same location when partitioning the dataset for training and testing. The data contains an average of 3-6 locations per class in each city, with up to 11 for some; there is no data for some scene/city combinations due to local restrictions regarding video recording in public places. 
For each location, 2-3 sessions of few minutes each was recorded, to obtain 5-6 minutes of material. The original recordings were split into segments with a length of 10 seconds that are provided in individual files, each being characterized by the scene class, city, and recording location IDs. 
The audio content of TAU Urban Audio-Visual Scenes 2021 is a  34 h subset of the TAU Urban Acoustic Scenes 2019 dataset (40 h), because not all recordings from TAU Urban Acoustic Scenes 2019 had a corresponding video recording.  Statistics of TAU Urban Audio-Visual Scenes 2021 dataset are presented in Table \ref{tab:data-stats}.

Besides the consistent quality of both audio and video, a specific characteristic of this dataset is the fixed recording position: the person performing the recordings was instructed not to move or speak during the recording. This results in a binaural audio recording in which the movement of the sound sources is only related to their physical movement with respect to the microphones, and not influenced by motions of the head. The camera was mounted at chest level on the strap of the backpack, therefore the captured video contains a number of objects moving across a static background, with no interference from body or hand movements. 
This provides a very particular setup, consistent throughout the entire data. Even though our current study is only concerned with scene classification, we expect to see different uses of this dataset in the future, for example audio-visual sound event detection or sound localization by learning from the relationship between objects and sound sources in the scene. 

For the public release of the dataset we performed additional postprocessing for blurring the faces and car licence plates visible in the video clips. The dataset is publicly available. A training/test setup with approximately 70\% of the data included in the training set and 30\% in the test set is provided along with the data to help reproduce and compare the results of different studies.\footnote{TAU Urban Audio Visual Scenes 2021, Development dataset, \\ DOI 10.5281/zenodo.4477542}  

\begin{table}[]
    \small
    \centering
    \begin{tabular}{l|ccc}
    \toprule
    Scene & Files & Locations & Duration (min) \\
    \midrule
    Airport         & 978   & 29 & 163.0 \\
    Bus             & 1134  & 53 & 189.0 \\
    Metro           & 1065  & 53 & 177.5 \\
    Metro station   & 1279  & 46 & 213.2 \\
    Park            & 1392  & 40 & 232.0 \\
    Public square   & 1369  & 41 & 228.2 \\
    Shopping mall   & 1228  & 31 & 204.7 \\
    Street pedestrian & 1389 & 45 & 231.5 \\
    Street traffic  & 1387  & 42 & 231.2 \\
    Tram            & 1071  & 51 & 178.5 \\
    \midrule
    Total           & 12292 & 431 & 2048.7 \\
   \bottomrule
   \end{tabular}
    \caption{TAU Urban Audio-Visual Scenes 2021 statistics}
    \label{tab:data-stats}
\end{table}

\section{Case study on scene classification}
 \label{sec:avsc}
 
Classification tasks are common in both audio and visual domains, with specific target classes such as objects (in image or video), sound events (in audio), and human activities (in audio or video) or multimedia events (in video). 
 
\subsection{Acoustic and visual scene classification}

Scene classification is defined as the task of assigning to the input audio clip or image a semantic label that characterizes the environment represented by the data \cite{barchiesi2015acoustic, Xie2020}. 
These labels are defined by humans and represent the general characteristics of the environment, for example airport, park, street, etc. 
This task is relevant for example in context-aware devices which change the operation mode of a device based on the context. 

Acoustic scene classification (ASC) is based on the analysis of the audio signal recorded at the scene, under the assumption that the general acoustic characteristics of the scenes are recognizable. State-of-the-art solutions are based on spectral features, most commonly the log-mel spectrogram, and convolutional neural network (CNN) architectures, often used in large ensembles \cite{Mesaros2019}. 
In comparison, visual scene classification (VSC) from images has a longer history and more types of approaches, e.g. global attribute descriptors, learning spatial layout patterns, discriminative region detection, and more recently hybrid deep models \cite{Xie2020}. 

\subsection{Audio-visual scene classification}
\label{sec:system}

Recent work in audio-visual analysis includes joint learning of audio and visual features, of which of special interest for our classification task is the $L^3$-Net architecture (look, listen and learn) \cite{DBLP:journals/corr/ArandjelovicZ17}. $L^3$-Net is trained for the audio-video correspondence (AVC) task, i.e. to predict if the audio and visual information presented at its input are related. The network contains two subnetworks, one for audio and another for video, each having four convolutional blocks which include two convolutional layers,  batch normalization and ReLu activation. The features outputted from each subnetwork are concatenated and fed into fusion layers that output the probabilities for the correspondence. For complete details on the network structure and training, we refer the reader to \cite{DBLP:journals/corr/ArandjelovicZ17}.

The OpenL3 network \cite{Cramer2019} is an open source implementation of $L^3$-Net, trained on AudioSet \cite{Gemmeke2017}. Authors of \cite{Cramer2019} provide an example of how to use the pretrained network on the downstream task of acoustic scene classification, by using audio embeddings generated by the audio branch, followed by feed-forward layers. The system was shown to outperform state of the art systems at the time on a few different datasets.

\begin{figure} 
    \centering
    \includegraphics[width=1.0\columnwidth]{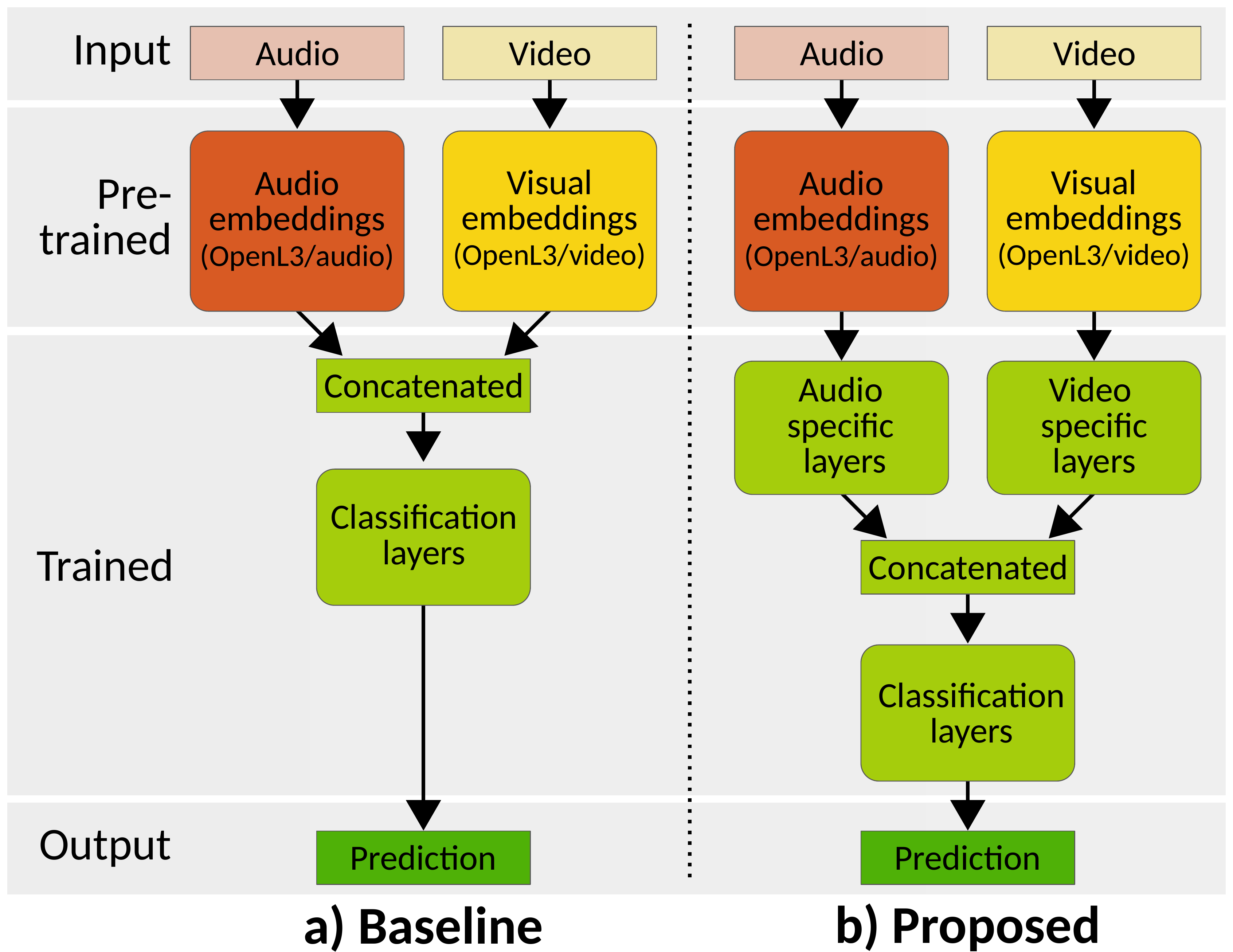}
    \caption{Neural network for audio-visual scene- recognition. a. Baseline: Audio and video embeddings are concatenated into a joint feature vector. b. Proposed: Separate audio and video networks first learn task-specific representations.}
    \label{fig:system}
    \vspace{-8pt}
\end{figure}

We propose a system for audio-visual scene classification (AVSC) based on OpenL3, using early fusion of the audio and video embeddings. 
Fig.~\ref{fig:system}.a shows the conventional approach for early fusion, where the audio and video embeddings are concatenated before being fed into the classification layers which are trained for the AVSC task. Our proposed system, presented in Fig.~\ref{fig:system}.b, uses separately trained ASC and VSC systems to generate more representative features for the task. 
Then their outputs at the second-last layer, used as feature representations for the AVSC task, are concatenated and fed into the classification layers. 
Classifiers of Fig.\ref{fig:system}.a contain three feed-forward layers of size 512, 128, and 64 and an output layer consisting of 10 neurons. Classifiers of Fig.\ref{fig:system}.b have two feed-forward layers of size 128 and 10.
The network in Fig.\ref{fig:system}.a is similar to the original OpenL3, but uses the representations learned from the AVC task to solve a classification problem. 

The hyperparameters of the signal processing and system training are the following. Audio embeddings are calculated with a window length of 1~s and a hop length of 0.1~s, extracted using the original OpenL3 network trained on the environmental AudioSet, using 256 mel filters. Each 1~s of audio is represented by an audio embedding vector of length 512. 
Video embeddings are calculated at a frame rate of 10 fps using the environmental videos. The frame corresponding to the center of each analysis window of the audio clip is resized such that the smallest size is 256, then the center 224x224 patch is given as input to the video subnetwork. Each 1~s segment of the video is represented by a video embedding vector of length 512. The audio and video embedding features are preprocessed using z-score normalization for bringing them to zero mean and unit variance. 

The system is implemented using PyTorch 
framework. The weights are updated using Adam optimizer \cite{kingma2014adam}, with a learning rate set to 0.0001 and weight decay of 0.0001. The models are trained for 200 epochs with a batch size of 64, using cross entropy loss. The models with best validation loss are retained. In the test stage, the system predicts an output for each 1~s segment of the data; the final decision for a clip is based on the maximum probability over 10 classes after summing up the probabilities that the system outputs for segments belonging to the same audio or video clip.

\section{Results}
\label{sec:results}

\begin{figure*} 
    \centering
    \includegraphics[scale=0.9]{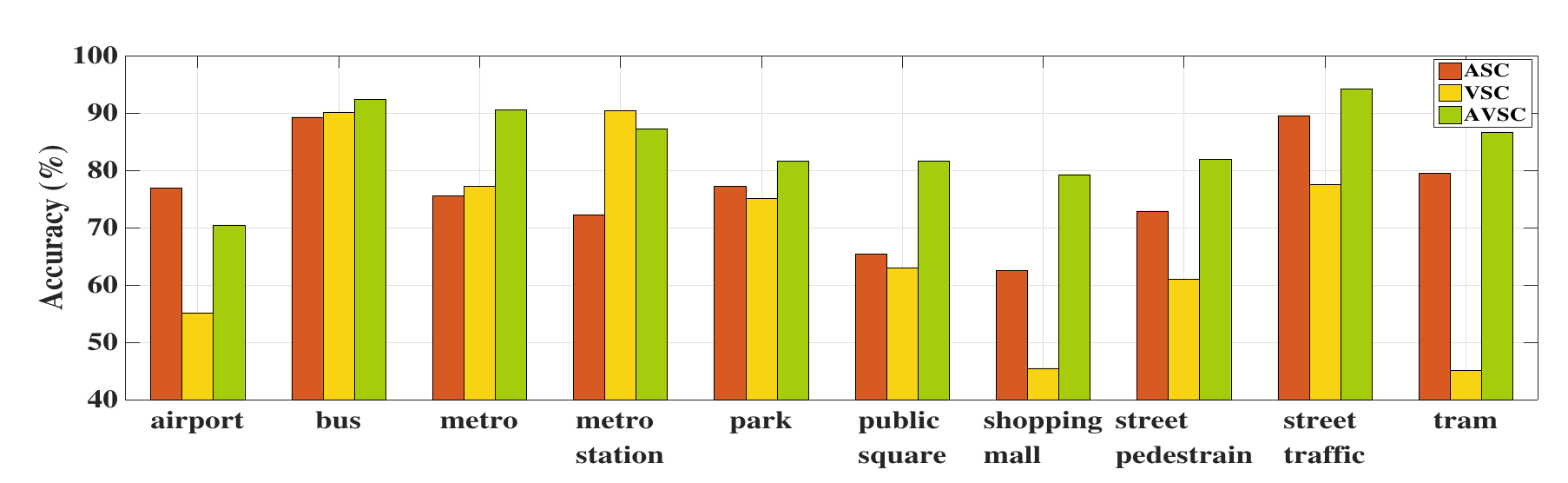}
    \caption{Class-wise performance for the OpenL3-based acoustic and visual scene classifiers, compared to the proposed method.}
    \label{fig:class-wise-avsc}
    \vspace{-8pt}
\end{figure*} 

The proposed system is evaluated against the separate acoustic scene classification and visual scene classification performance, to analyze the improvement brought by the joint model. We also compare the early fusion and late fusion alternatives of the two systems. 
Early fusion is achieved by concatenating the audio embeddings with the video embeddings before the classifier, as shown in  Fig.\ref{fig:system}.a, while late fusion is achieved by combining the outputs of separately trained audio and video models over ten classes, to obtain the final decision. The results are presented in Table \ref{tab:ex_results}. 

The accuracy of the separately trained ASC system based on OpenL3 embeddings is 75.8\%. We compare this with the McDonnel\_USA system \cite{Gao2019}, one of the top ranked submissions in the DCASE 2019 Challenge, which uses residual networks and log-mel spectrogram. Its performance on the TAU Audio-Visual Scenes 2021 data is 82.84\%, which is significantly higher than our OpenL3-based ASC system. 


The accuracy of the separately trained VSC system based on OpenL3 is 68.4\%, much lower than the performance of the audio side. We compare this to an approach based on ResNet50, a widely available pretrained network that has provided state-of-the-art results in various image classification tasks \cite{He_2016_CVPR}. We trained a ResNet50-based classifier using ten frames per video clip, extracted by randomly selecting one frame in each one second of the video clip. The system obtains an accuracy of 77.9\%, which is significantly higher than the performance of our OpenL3-based VSC system. 


We construct an audio-visual scene classifier as an ensemble of MCDonnell\_USA system (AC) and ResNet50 (VSC), using late fusion. Its overall accuracy 85.8\%. The late fusion of the OpenL3-based ASC and VSC classifiers provides a performance of 80.8\%. 

Early fusion of the OpenL3-based ASC and VSC systems outperforms significantly the individual subsystems, with the conventional early fusion (Fig.~\ref{fig:system}.a) obtaining 82.2\% accuracy, and the proposed method 84.8\%. The overall improvement obtained with the proposed method is significant compared to all individual systems. Class-wise performance of the proposed method in comparison with the single-modal systems is illustrated in Fig.~\ref{fig:class-wise-avsc}. We observe different class-wise behavior in the audio and visual domain, with the video models being more powerful for the metro station and metro classes, but having very low performance compared to the audio models in tram class (45.1\% compared to 79.5\%). The fusion helps improving the performance for some classes, with very high gains for shopping mall (17\% absolute improvement compared to audio only),  and street pedestrian (10\% absolute improvement). On the other hand, for some scenes the audio-visual system performs worse than the audio only system (6\% lower for airport).

\begin{table}[]
    \small
    \centering
    \begin{tabular}{l|c}
    \toprule
    Method & ACC  \\
    \midrule
    Audio only &  75.8\%   \\
    Video only &  68.4\%\\
    \midrule
    Early A-V fusion &  82.2 \%\\
    Late A-V fusion & 80.8 \%\\
    Proposed early A-V fusion & \textbf{84.8 \%} \\
    \bottomrule
    \end{tabular}
    \caption{Classification accuracy of the proposed method in comparison with the networks using audio only, video only, and early and late fusion. All systems are based on OpenL3.}
    \label{tab:ex_results}
\end{table}

The state of the art performance for scene classification on the comparable TAU Urban Acoustic Scenes 2019 dataset is 85\% \cite{Chen2019}, surpassing the performance obtained by our joint system. However, the system in \cite{Chen2019} is an ensemble of 7 subsystems, and has 48 million parameters, while our complete AVSC system has only 13.4 million total parameters. In comparison, the McDonnell\_USA we used as baseline has only 3 million parameters, while ResNet50 has approximately 23 million trainable parameters, making our solution the lightest one computationally.  

\section{Conclusions and future work}

\label{sec:concl}
In this paper we introduced the first curated dataset for audio-visual scene analysis, and a case study of audio-visual scene classification. We have shown that joint learning of scene characteristics using the audio and visual domain signals brings significant improvement in performance compared to modeling the domains individually. 
The performance of our approach on the separate audio and video signals is slightly lower than dedicated state of the art systems using the same signal. However, our system has the advantage of using similar feature representation for audio and video, obtained within the same structure, and of being computationally less burdensome. It should be noted that the joint model does not always have an advantage compared to the separate acoustic or visual models.
The dataset is provided to the research community as open data with the hope that it will support novel directions of research in audio-visual scene analysis. The special characteristics of the data support object/source tracking, and the possibility for inter-modal knowledge transfer when one modality is not available.
\bibliographystyle{IEEEbib}
\bibliography{refs}

\end{document}